\documentclass[
reprint,
amsmath,amssymb,
prb,
]{revtex4-2}
\usepackage[utf8]{inputenc}
\usepackage{tikz}
\usepackage[export]{adjustbox}
\usepackage{ulem}
\usepackage{physics}
\usepackage{graphicx,graphics}% Include figure files
\usepackage{dcolumn}% Align table columns on decimal point
\usepackage{bm}% bold math
\usepackage{color}
\usepackage{lipsum}
\usepackage[percent]{overpic}
\usepackage{array}
\usepackage{nicefrac}
\usepackage{multirow}
\usepackage{latexsym,verbatim}
\usepackage{amsmath,amssymb,amsfonts}
\usepackage[breaklinks=true,colorlinks,citecolor=blue,linkcolor=blue,urlcolor=blue]{hyperref}
\makeatletter
\renewcommand{\thefigure}{\arabic{figure}}       % Make \ref{fig:...} output 1, 2, ...
\renewcommand{\fnum@figure}{Figure~\thefigure}   % Make caption show "Figure 1"
\makeatother

\begin{document}

\newcommand{\Ts}{T_{\rm s}}
\newcommand{\Tg}{T_{\rm g}}
\newcommand{\Tb}{T_{\rm b}}
\newcommand{\Qgis}{Q_{\rm GIS}}
\newcommand{\R}{\mathcal{R}}

\title{Out-of-equilibrium nonlinear model of thermoelectricity in superconducting tunnel junctions}
\author{Leonardo Lucchesi}
\affiliation{INFN Sezione di Pisa, Largo Bruno Pontecorvo 3, 56127 Pisa, PI, Italy}
\author{Federico Paolucci}
\affiliation{INFN Sezione di Pisa, Largo Bruno Pontecorvo 3, 56127 Pisa, PI, Italy}
\affiliation{Dipartimento di Fisica, Università di Pisa, Largo Bruno Pontecorvo 3, 56127 Pisa, PI, Italy}

\begin{abstract}
    Thermoelectricity in superconducting tunnel junctions has always been studied under the hypothesis of equilibrium between the cold side and the thermal bath, usually in the linear regime. We define a more complete out-of-equilibrium nonlinear numerical model that reduces to the equilibrium linear model in the low-power limit. We find that the linear model does not correctly describe the behavior of superconducting tunnel junctions for parameters that are reasonable in practical experimental setups. Subsequently, we present the qualitative and quantitative differences between the models, discovering that for high power, the junction saturates and then inverts its behavior. Finally, we also clarify the difference between linear and nonlinear thermoelectricity and devise a new criterion to find nonlinear thermoelectricity in the parameter space. 
\end{abstract}
\maketitle
\section{Introduction}
The idea of studying thermoelectricity in nanometric superconducting junctions may sound baffling until one considers that
heat is an inevitable byproduct of information processing \cite{Shannon1948}, even at cryogenic temperatures. The development of solid-state quantum nanotechnologies such as qubits \cite{Raussendorf2001,Koch2007,Valenzuela2006,Kjaergaard2020,Burkard2023} or transition-edge sensors (TES) \cite{Irwin1995,Ade2015,Crowley2018,Mairs2021} requires a consistent heat management design, whose difficulty strongly increases with the number of active elements (qubits or pixels). Qubit operation requires strict control of temperature \cite{Valenzuela2006}, and working thermoelectric elements could be extensively used for control thermometry \cite{Giazotto2015} and detection of radiation and particles \cite{Heikkilae2018}. On its own, the study of heat transport at the nanoscale shows interesting phenomena caused by the coherence of the heat carriers, for example, heat quantization \cite{Schwab2000,Jezouin2013}, interferometry \cite{Giazotto2012} and diffraction \cite{JoseMartinezPerez2014}.
A related concept in the computing field is caloritronics \cite{Fornieri2017}, the idea of making friends with your enemy by processing heat signals instead of electronic signals. Currently, designs for heat diodes \cite{Chang2006,MartinezPerez2015}, splitters \cite{Timossi2018}, valves \cite{Strambini2014}, transistors \cite{Paolucci2017}, logic gates \cite{Paolucci2018} and memories \cite{Ligato2022} have been realized, and in this framework, thermoelectric modules could act as heat-electricity transducers. Another potential application of cryogenic thermoelectricity is the ultrasensitive passive detection of radiation \cite{Giazotto2006}, which differs from standard detectors \cite{Pirro2017} that require a current or voltage bias \cite{Irwin1995,DeLucia2024}. \\
For all these applications, we would need a thermoelectric module working at cryogenic temperatures, but bringing standard thermoelectric materials under 1 Kelvin causes a strong reduction in the Seebeck coefficient \cite{Mao2020,Chung2000,Sidorenko2019}, and using 3D semiconductors is inefficient because of temperature-induced charge depletion in the bands and a zero density of states at the band edge \cite{Grosso2014}. Reducing the dimensionality could solve the problem \cite{Hicks1993,Hicks1993a}, and we see that quantum dot setups show good thermoelectric efficiencies \cite{Sothmann2014,Jaliel2019,Urban2015} but face important scaling issues \cite{Urban2015}. In recent years, superconducting hybrid structures have been introduced as a setup for cryogenic thermoelectricity \cite{Ozaeta2014,Marchegiani2020}. In superconducting tunnel junctions, alignment between two different densities of states (DOS) allows the creation of different transport regimes that can induce voltage-controlled cooling \cite{Melton1980,Nahum1994,Giazotto2006,Leivo1996,Tarasov2020,Manninen1999}, and negative absolute conductance induced by a temperature gradient \cite{Kolenda2016,Germanese2022,Germanese2023}, which is the hallmark of thermoelectricity generation. 
Two main mechanisms produce thermoelectricity in these junctions: a linear mechanism inducing a thermoelectric voltage that can be linearly expanded in the temperature gradient $\delta V\propto \delta T$ \cite{Ozaeta2014}, and a nonlinear mechanism producing a thermovoltage $V=\pm |\Delta(T)-\Delta^\prime(T^\prime)|$ which just depends on the superconducting gap of the two materials $\Delta^{(\prime)}(T)$ and is almost independent of the thermal gradient for $T\ll T_c$ \cite{Marchegiani2020}, where $T_c$ is the superconductor critical temperature. Both mechanisms offer alluring pathways to produce devices based on superconducting hybrid junctions such as thermometers \cite{Giazotto2015,Vischi2020}, heat engines \cite{Ozaeta2014,Germanese2022}, and detectors \cite{Vischi2020,Heikkilae2018}. Graphene-Insulator-Superconductor (GIS) junctions host both mechanisms because of the linear DOS of graphene \cite{Bianco2024}. For superconducting junctions in the tunneling regime, thermal dynamics %and noise have 
has been extensively studied in the linear regime of the thermal dynamics equations \cite{Mather1982,Richards1994,Golubev2001,Chakraborty2018} around a working point dictated by a voltage or current bias \cite{Golubev2001}. However, as already noticed for quantum dot-based setups, thermal gradients easily bring nanostructures out of the linear regime of energy balance equations \cite{Whitney2013,Sothmann2014,Talbo2017}.\\
In this article, we study the nonlinear regime in superconducting hybrid tunnel junctions via numerical simulations of the thermal balance and current equations. To the best of our knowledge, all works on superconducting tunnel junctions assume equilibrium transport with one side fixed at the thermal bath temperature $\Tb$ \cite{Heikkilae2018}. 
For the first time in the tunneling regime, we utilize the full nonlinear model to describe an out-of-equilibrium state where the colder temperature can be different from $\Tb$. We use a GIS junction as a toy model, the simplest and most general superconductor hybrid junction that exhibits both linear and nonlinear thermoelectricity; however, our results apply to most thermoelectric superconducting junction setups.\\
We will begin in Section \ref{sec:model} by introducing the full nonlinear model for the junction and showing how the linear model is obtained.\\
In Section \ref{sec:failures}, we will identify how the linear model fails to correctly describe the physics of the junction, even for relatively low impinging powers and for any nonzero subgap state density. Also, we will introduce a new criterion for identifying the onset of nonlinear thermoelectricity, another phenomenon not described by linear modelling. \\
Finally, Section \ref{sec:effects} will focus on the effects of nonlinearity on the observable quantities.  

\section{Model}\label{sec:model}
\begin{figure}[tb]
\centering
\includegraphics[width=\columnwidth]{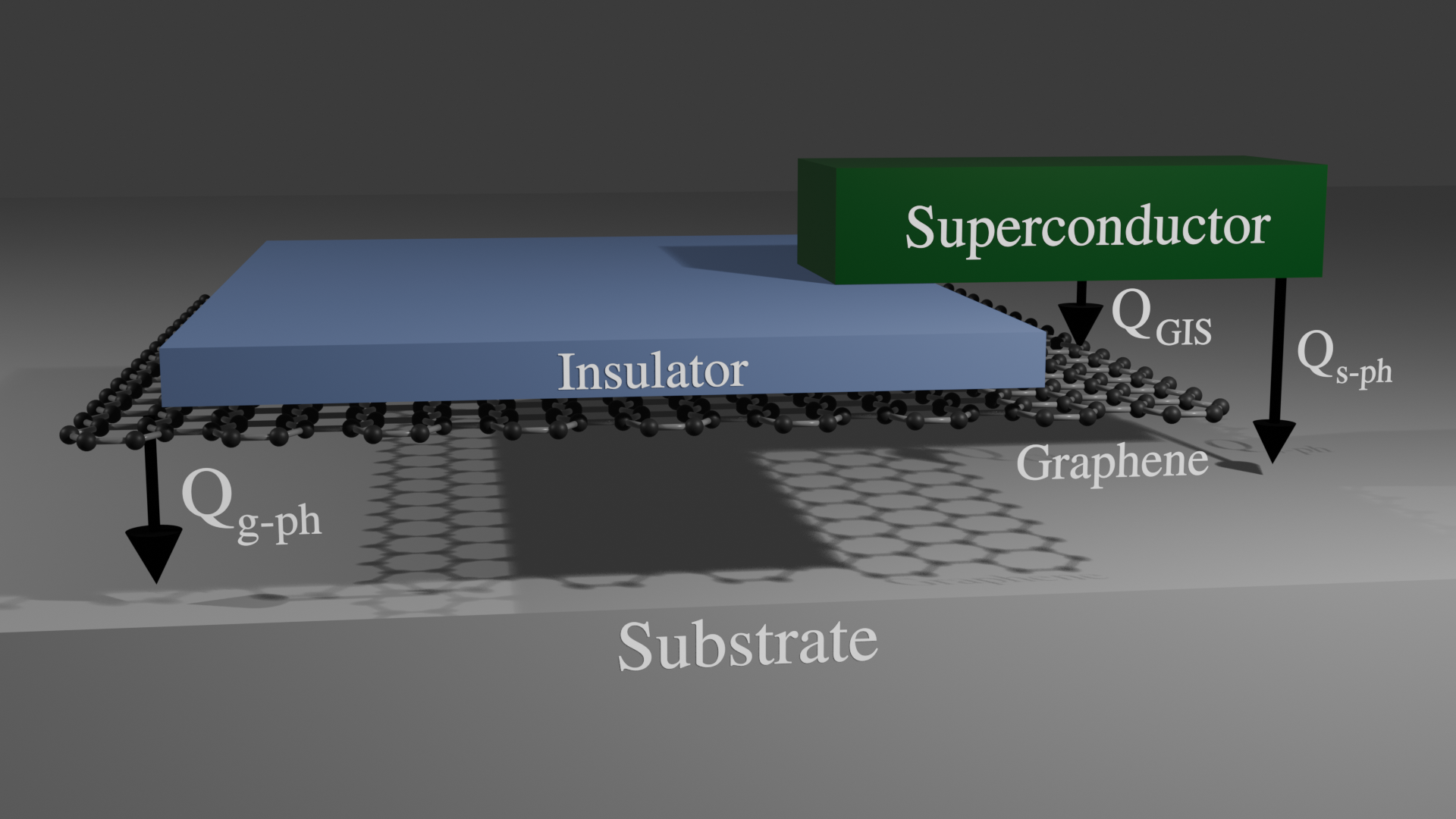}
\caption{Schematic depiction of GIS junction showing all the building blocks and heat flows between them. 
}
\label{fig2leads}
\end{figure}
To understand the effect of linearization schemes and out-of-equilibrium conditions, we write down the full nonlinear model of thermoelectricity of a junction. Essentially, the GIS junction produces thermoelectricity when the Fermi energy of graphene $E_F\neq 0$ because there is a difference in the graphene density of states (DOS) $n_{\rm g}(E)$ at energies $+\Delta_0\equiv\Delta(\Ts=0)$ and $-\Delta_0$ being populated by the Fermi functions $f(E-eV,\Tg)$ and $f(E,\Ts)$, with $\Ts$ being the temperature of the superconductor, $\Tg$ being the temperature of graphene and $V$ the potential applied to graphene. This $n_{\rm g}(\Delta_0)-n_{\rm g}(-\Delta_0)$ difference induces an unbalance between the tunneling currents from superconductor to graphene and vice versa $I_{\rm sg}\neq I_{\rm gs}$ if $\Ts\neq \Tg$.
When $\Tg>\Ts$, $V$ shifts in energy the hot distribution $f(E-eV,\Tg)$ with respect to the superconductor DOS $n_{\rm s}(E,\Ts)$ (defined later in Eq. \ref{eq:supdos}). This means that the Seebeck potential is linear in temperature $\delta V_S\sim \delta T$, as the effect of a $V_S\ll \Delta_0/e$ is able to compensate the Peltier current $I_0$ induced by $\delta T$. When $\Ts>\Tg$, only the cold distribution is shifted, so a potential $V_S\sim \Delta_0/e$ is needed to have $I=0$ through the junction.\\
With this picture in mind, we can write the full thermal model and electric model of the junction, leaving both temperatures free to differ from $\Tb$.
\subsection{Full model}
We can describe the thermal dynamics of the junction by using two coupled thermal balance equations for the superconductor $T_s$ and the graphene $T_g$ temperatures,
\begin{equation}\label{eq:thermdyn}
    \begin{cases}
      \displaystyle C_{\rm s}(T_{\rm s}) \frac{d\Ts}{dt}=-Q_{\rm GIS}(T_{\rm s},T_{\rm g})-Q_{\rm s-ph}(T_{\rm s})+P_{\rm s} \\[8pt]
      \displaystyle C_{\rm g}(T_{\rm g}) \frac{dT_{\rm g}}{dt}=Q_{\rm GIS}(T_{\rm s},T_{\rm g})-Q_{\rm g-ph}(T_{\rm g})+P_{\rm g}, 
    \end{cases}
\end{equation}
where $C_{\rm s,g}$ are the thermal capacitances of the superconductor and the graphene, $Q_{\rm s-ph}, Q_{\rm g-ph}$ and $Q_{\rm GIS}$ are the heat flows depicted in Fig. \ref{fig2leads} and respectively defined in Eq.\ref{eq:epsc},\ref{eq:qdirtycleang} and \ref{eq:qgis}, and $P_{\rm s,g}$ are external radiant powers impinging on the superconductor and the graphene. From now on, we will only consider power impinging on the superconductor $P_{\rm g}=0$, because with $\Tg>\Ts$ we only have linear thermoelectricity \cite{Bianco2024}.\\
The thermal capacitance of a superconductor with quasiparticle DOS $n_{\rm s}(E,T)$ can be obtained from the entropy of the quasiparticle gas $S_{\rm s}(T)$ by \cite{Paolucci2023}
\begin{subequations}
\begin{align}\label{eq:thermsuper}
    C_{\rm s}(T)&=T\frac{dS_{\rm s}}{dT}\\
    S_{\rm s}(T)&=-2\mathcal{V}_{\rm s} N_F k_B \int dE\;[n_{\rm s}f\ln{f}](E,T)\\
    n_{\rm s}(E,T)&=\frac{(E+i\Gamma)}{\sqrt{(E+i\Gamma)^2-\Delta^2(T)}},\label{eq:supdos}
\end{align}
\end{subequations}
where $k_B$ is the Boltzmann constant, $\mathcal{V}_{\rm s}$ is the volume of the superconductor, $N_F$ is the DOS at the Fermi energy $E_F$ for the metal phase of the superconductor, and $\Gamma$ is the Dynes parameter \cite{Dynes1984} accounting for the DOS induced by impurities and interfaces in the middle of the superconducting gap (see Sec.\ref{sec:gamma} for further discussion).\\
Since the characteristic thermal energy is much smaller than the Fermi energy $k_B \Tg\ll E_F$, we can approximate its thermal capacitance at low temperatures by using the specific heat of a Fermi liquid obtained from a Sommerfeld expansion and by approximating the graphene DOS $n_{\rm g}(E)\approx n_{\rm g}(E_F)$, obtaining
\begin{equation}
    C_{\rm g}(T)=A_{\rm g}\gamma T \qquad \qquad \gamma=\frac{\pi^2}{3}k_B^2 n_{\rm g}(E_F),
\end{equation}
where $A_{\rm g}$ is the graphene area, and $n_{\rm g}(E)$ is analytically computed from the nearest neighbor model \cite{CastroNeto2009}

\begin{subequations}\label{eq:graphenedos}
	\begin{align}
		n_{\rm g}(E) &= \frac{4}{\pi^2} \frac{\lvert E \rvert}{t^2} \frac{1}{\sqrt{Z_0}}
\mathbf{F}\left(\frac{\pi}{2}, \sqrt{\frac{Z_1}{Z_0}}\right) \\
		Z_0 &= \left(1\!+\!\Big\lvert
\frac{E}{t}\Big\rvert\right)^2\!-\!\frac{\left[\left(\frac{E}{t}\right)^2\!-\!1\right]^2}{4} \\
		Z_1 &= 4\Big\lvert \frac{E}{t}\Big\rvert 
	\end{align}
\end{subequations}
with $t=2.7\,{\rm eV}$ being the nearest neighbor hopping parameter. $E_F$ is defined by inverting the equation
\begin{equation}
    n=\int_{-\infty}^{\infty} dE\,n_{\rm g}(E)f(E-E_F).
\end{equation}
We described the phonon-induced heat dissipation in the superconductor $Q_{\rm s-ph}$ via a nonequilibrium approach within the quasiclassical approximation \cite{Timofeev2009}, assuming that the heat exchange between phonons in the superconductor and phonons in the substrate is so efficient that we can assume vanishing Kapitza resistance \cite{Giazotto2006}, i.e. they are at equilibrium with the bath temperature $T_{\rm b}$. We can find the heat flow from \cite{Timofeev2009}
\begin{multline}\label{eq:epsc}
Q_{\rm s-ph} = -\frac{\Sigma
\mathcal{V}_{\rm s}}{96\zeta(5)k_B^5}\int_{-\infty}^{\infty} dE\, E
\int_{-\infty}^{\infty} d\epsilon \,\epsilon^2{\rm
sign}(\epsilon)M_{E,E+\epsilon}\\
\times \left[\coth\left(\frac{\epsilon}{2k_BT_{\rm
b}}\right)(f_E^{(1)}-f_{E+\epsilon}^{(1)})-f_E^{(1)}f_{E+\epsilon}^{(1)}+1\right],
\end{multline}
where $f_E^{(1)}=f_0(-E,T_{\rm s})-f_0(E,T_{\rm s})$, and
%$M_{E,E+\epsilon}$ is given by 
$M_{E,E'}=n_{\rm s}(E)n_{\rm s}(E')[1-\frac{\Delta^2(T_{\rm s})}{EE'}]$.\\
Heat dissipated by phonons in graphene $Q_{\rm g-ph}$ is described by two power law formulas, respectively in the clean regime $ql_{\rm mfp}\gg 1$ and the dirty regime $ql_{\rm mfp}\ll 1$ limits with $q$ being the phonon wavenumber \cite{Chen2012,Zihlmann2019}

\begin{subequations}\label{eq:qdirtycleang}
\begin{align}
     Q_{\rm g-ph, clean} &= A_g \frac{\pi^{5/2} D_p^2 \sqrt{n} k_B^4}{15 \rho_M \hbar^4 v_F^2 c_s^3} \left(T_{\rm g}^4-T_{\rm b}^4\right) \\
       Q_{\rm g-ph, dirty} &= A_g \frac{2 \zeta(3) D_p^2 \sqrt{n} k_B^3}{\pi^{3/2} \rho_M \hbar^3 v_F^2 c_s^2 l_{\rm mfp}} \left(T_{\rm g}^3-T_{\rm b}^3\right),
\end{align}
\end{subequations}
where we have the graphene area $A_g$, the sound speed $c_s=2\cdot 10^4 \,{\rm m/s}$, the mass density $\rho_M=7.6\cdot 10^{-7}\,{\rm kg/m^2}$, the deformation potential $D_p=13\cdot e\,{\rm J}$, and the mean free path $l_{\rm mfp}=60\,{\rm nm}$. These formulas were obtained for temperatures much smaller than the Bloch-Gr{\"u}neisen temperature $T_{\rm BG}\sim 50\,{\rm K}$ by using a full nonequilibrium approach.\\
Finally, we describe the tunneling part by writing heat flow across the GIS junction as \cite{Vischi2020}
\begin{multline}\label{eq:qgis}
    Q_{\rm GIS}(T_{\rm s},T_{\rm g},V)=\frac{1}{e^2R_t}\int dE \; (E-eV)\rho_{\rm g}(E-E_F-eV)\cdot\\ \cdot \rho_{\rm s}(E,T_{\rm s}) \left[f(E,T_{\rm s})-f(E-eV,T_{\rm g})\right],
\end{multline}

where $\rho_{\rm i}=n_{\rm i}/n_{\rm i}(E_F)$ are the densities of states normalized over their value at the Fermi energy (for the superconductor it is the value in the metallic state) $R_t$ is the tunneling resistance, containing information on the geometry of the junction $R_t\propto (A_j n_{\rm g}(E_F)n_{\rm s}(E_F)\lvert T\rvert^2)^{-1}$ where $A_j$ is the area of the junction and $\lvert T\rvert^2$ is an energy independent transmission coefficient.\\ 
For the sake of this article, we can use the simplest possible model for electric transport, described by the tunneling integral
\begin{multline}\label{eq:iv}
    I(V,\Ts,\Tg)=\frac{1}{eR_T}\int\,dE\; \rho_{\rm s}(E,\Ts)\cdot \\
    \cdot \rho_{\rm g}(E-E_F-eV)[f(E-eV,\Tg)-f(E,\Ts)],
\end{multline}
and restricting the description to two setups:
\begin{itemize}
    \item Open circuit: We connect an ideal voltmeter to the junction, thus no current can flow in the circuit. We measure the Seebeck voltage $V_S$.
    \item Closed circuit: We connect an ideal ammeter to the junction, thus no voltage difference can form across the junction. We measure the Peltier current $I_0$.
\end{itemize}
\subsection{Linear model}
Since the full system of differential equations describing transport in the junction is not analytically solvable, one can linearize the model to obtain approximate results by considering small temperature and potential differences.
In \cite{Ozaeta2014}, the authors expand in $\delta T= (\Ts -\Tg)$ and then compute the linear transport coefficients using the average temperature $T=(\Ts -\Tg)/2$.  This linear approximation is equivalent to expanding the electrical and the heat currents in a small temperature difference between the superconductor and the bath $\delta \Ts\equiv \Ts-\Tb$ and keeping $\Tg=\Tb$, because if $\Tg$ is not close to $\Tb$, $Q_{\rm g-ph}\gg Q_{\rm GIS}$ in any realistic setup, enforcing $\Tg=\Tb$. We use an analogous equivalence by expanding in the applied potential $\delta V\equiv V$. 
We obtain
\begin{equation} \label{eq:lintrans}
\displaystyle{
    \begin{pmatrix} I\\ Q
    \end{pmatrix}=
    \begin{pmatrix}
        G & \alpha \\
        \alpha & G_{\rm th}\Tb
    \end{pmatrix}\cdot
    \begin{pmatrix}
        \delta V \\ \frac{\delta T}{\Tb}
    \end{pmatrix}}
\end{equation}
with the linear transport coefficients defined as
\begin{subequations}\label{eq:lintranscoeff}
\begin{align}
    G \equiv \frac{\partial I}{\partial V} &= \frac{1}{R_t}\int dE\, \frac{\rho_{\rm g}(E-E_F)\rho_{\rm s}(E,\Tb)}{4k_B\Tb\cosh^2(\frac{E}{2k_B\Tb})}\\
    \alpha \equiv \frac{\partial Q_{\rm GIS}}{\partial V} &= \frac{1}{eR_t}\int dE\, E\frac{\rho_{\rm g}(E-E_F)\rho_{\rm s}(E,\Tb)}{4k_B\Tb\cosh^2(\frac{E}{2k_B\Tb})}\\
    G_{\rm th} \equiv \frac{\partial Q_{\rm GIS}}{\partial T}  &= \frac{1}{e^2R_t}\int dE\, E^2\frac{\rho_{\rm g}(E-E_F)\rho_{\rm s}(E,\Tb)}{4k_B\Tb\cosh^2(\frac{E}{2k_B\Tb})}.
\end{align}
\end{subequations}
The linear transport matrix in Eq. \ref{eq:lintrans} is symmetric because of the Onsager relation $\partial Q/\partial V=\partial I/\partial T$ \cite{Onsager1931}, which is valid as long as time reversal symmetry is not broken in the quantum propagation processes, for example, by a magnetic field. \\
The open circuit setup imposes $I=0$, and a Seebeck potential $V_S=-(\alpha/G\Tb)\delta T$ forms to cancel the thermocurrent generated by $\delta T$. Using a linearized version of the first thermal balance equation in Eq.\ref{eq:thermdyn}, and the linear approximation for $Q$ in Eq.\ref{eq:lintrans}, we can compute the temperature difference induced by $P_{\rm s}$ as
\begin{equation}\label{eqdeltaTlin}
	\delta T=\frac{G\Tb}{G^{\rm tot}_{\rm th}G\Tb-\alpha^2}P_{\rm s},
\end{equation}
where $G^{\rm tot}_{\rm th}=G_{\rm th}+G_{\rm s-ph}$ with $G_{\rm s-ph}\equiv \partial Q_{\rm s-ph}/\partial \Ts|_{\Tb}$. 
\section{Failures of the linear model}\label{sec:failures}
\subsection{Limits of linear model}\label{sec:limlin}
\begin{figure}[tb]
\includegraphics[width=\columnwidth]{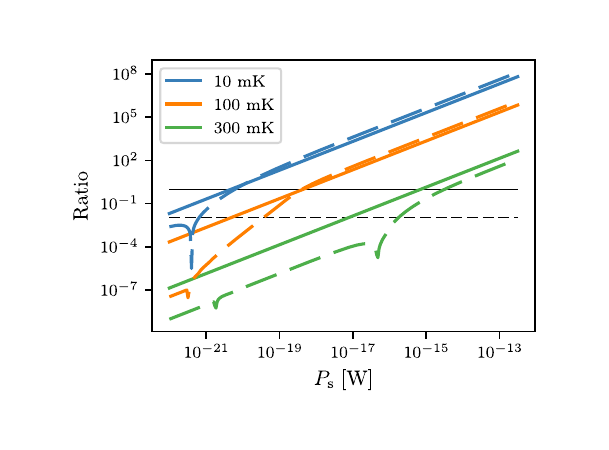}

\caption{Linearity check for thermal dynamics in realistic setups (text for details). Solid lines: Impinging power $P_{\rm s}$ dependence of ratio $r_{\rm lim}\equiv\delta T/\delta T_{\rm lim}$ of linear model over linear limit. The ratio becomes larger than $1$ (black line) for impinging powers that are too small for realistic applications, forbidding the use of the linear model for larger powers. Dashed lines: $P_{\rm s}$ dependence of relative difference $r_{\rm nl}\equiv|\delta T-\delta T_{\rm nl}|/T_{\rm b}$ between linear and nonlinear model. This second linearity criterion confirms the validity of the first one even if the first one is computed within the linear model. Parameters in main text.}
\label{fig2}
\end{figure}
For the linear approximation to hold, first-order terms must be much larger than second-order terms, implying
\begin{equation}\label{eqdeltaTlim}
    \delta T \ll 2\left|\frac{\partial Q_{\rm GIS}/\partial \Ts + \partial Q_{\rm s-ph}/\partial \Ts}{\partial^2 Q_{\rm GIS}/\partial \Ts^2 + \partial^2 Q_{\rm s-ph}/\partial \Ts^2}\right|\equiv \delta T_{\rm lim}.
\end{equation}
Therefore, we can assess the validity of the linear model by comparing the linearly computed $\delta T$ from Eq.\ref{eqdeltaTlin} with the limit calculated in Eq.\ref{eqdeltaTlim} for $P_{\rm s}$ typical for different uses such as nanodetection \cite{DeLucia2024} and cryogenic heat engines \cite{Germanese2022}. Their ratio $r_{\rm lim}\equiv\delta T/\delta T_{\rm lim}$ is represented in Fig.\ref{fig2} for different bath temperatures using solid lines and material parameters reasonable for a nanodetector: aluminum with $\Delta_0=2\cdot 10^{-4}\,{\rm eV}$, volume $\mathcal{V}_{\rm s}=10^{-19}\,{\rm m}^3$, $\Gamma=10^{-4}\Delta_0$, junction area $A_j=10^{-2}\,\mu{\rm m}^2$ and graphene area $A_{\rm g}=30\,\mu{\rm m}^2$. For control purposes, we also represented the ratio $r_{\rm nl}\equiv|\delta T-\delta T_{\rm nl}|/\Tb$, comparing the temperature difference obtained from the full model $\delta T_{\rm nl}$ with the linearly computed one. In Fig.\ref{fig2}, we see that the criterion for linearity obtained from within the linear model describes the power at which the linear model breaks down with reasonable precision. In Fig.\ref{fig2}, we can see that linearity is broken for different $P_{\rm s}$, $\sim 10^{-20} \,{\rm W}$ for $\Tb=10 \,{\rm mK}$, $\sim 10^{-18}\,{\rm W}$ for $\Tb=100 \,{\rm mK}$ and $\sim 10^{-15} \,{\rm W}$ for $\Tb=300\, {\rm mK}$. While $r_{\rm lin}$ trivially shows a linear behavior described in Eq.\ref{eqdeltaTlin}, $r_{\rm nl}$ depends on the results of the nonlinear model. For low $P_{\rm s}$, $r_{\rm nl}$ becomes negligible as expected, and its sign has little meaning. For all the considered bath temperatures $\delta T_{\rm nl}$ becomes larger than $\delta T$ below a certain value of $P_{\rm s}$, but $\delta T$ quickly becomes larger than $\delta T_{\rm nl}$ with larger $P_{\rm s}$ because $\delta T_{\rm nl}$ is contained by the superlinearly increasing $Q_{\rm s-ph}$ and $Q_{\rm g-ph}$ in the two materials. When $r_{\rm nl}>1$, $\delta T$ is dominating, inducing the same slope as $r_{\rm lim}$.\\
Naturally, device geometry affects the power $P_1$ defined by $r_{\rm nl}(P_1)>10^{-2}$, i.e., when the linear model ceases to be an accurate description of our system. The linear temperature difference depends on the junction area $\delta T\sim P_{\rm s}/A_j$ via the tunnel resistance $R_t$, while the linearity threshold $\delta T_{\rm lim}$ depends on $A_j$ via the derivatives of $\partial Q_{\rm GIS}/\partial (\Ts,\Tg)\propto A_j$, and on the superconductor volume $\mathcal{V}_{\rm s}$ via $\partial Q_{\rm s-ph}/\partial \Ts\propto \mathcal{V}_{\rm s}$. Both $\delta T$ and $\delta T_{\rm lim}$ are dominated by the derivatives of $\partial Q_{\rm GIS}/\partial (\Ts,\Tg)$, especially for lower $\Tb$. Therefore, the dependence of $P_{1}$ on $A_j$ and $\mathcal{V}_{\rm s}$ can be reasonably represented by a linear dependence on the junction area alone $P_1\sim A_j$. This tells us that the possible range of variation for $P_1$ corresponds to the possible range of variation for $A_j$, which can reasonably vary between $\sim 10^{-2}\mu{\rm m}^2$ and $\sim 1\mu{\rm m}^2$. Therefore, geometry alone can make $P_1$ vary by two orders of magnitude at most, confirming that linearity breaking happens at very low $P_{\rm s}$ for any reasonable geometry.\\
These considerations motivate the development of a nonlinear model, which will be discussed in the following section, Sec. \ref{sec:effects}.
\subsection{Subgap states}\label{sec:subgap}
\begin{figure}[tb]
\includegraphics[width=\columnwidth]{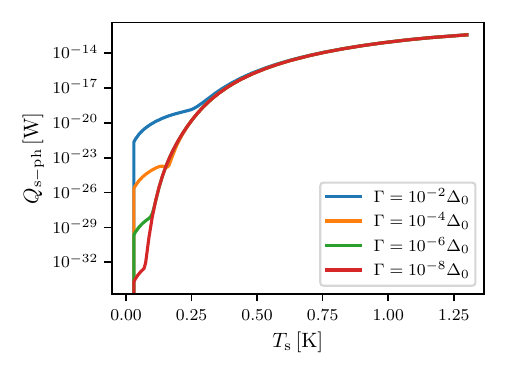}
\caption{Effect of subgap states on superconductor dissipation. The dependence of $Q_{\rm s-ph}$ on temperature for different values of the Dynes parameter $\Gamma$ shows the effect of non-idealities on heat dissipation. Effects are non-negligible only for the lowest temperatures, for which $n_{\rm s}(\Delta)f(\Delta)\lesssim\Gamma$.}
\label{fig3}
\end{figure}

The Dynes parameter $\Gamma$\cite{Dynes1984} is a standard phenomenological approach for incorporating nonidealities of real superconductors into simplified models of densities of states \cite{Giazotto2006}. $\Gamma$ is essential to consider the effects of non-idealities, as they are important for applications and can significantly alter the physics. Intuitively, $\Gamma/\Delta$ represents the number of states at the Fermi energy in a superconductor, as one can see from Eq.\ref{eq:thermsuper} in the limit $E\rightarrow 0$. In Fig.\ref{fig3}, we show the effects of using different values for $\Gamma$ when computing $Q_{\rm s-ph}$. 
For lower temperatures, we observe that larger $\Gamma$ implies greater dissipation, an intuitive consequence of having states that can be occupied by quasiparticles, which can then interact with phonons. For larger temperatures, the effect of $\Gamma$ is dominated by quasiparticles with $E>\Delta$. \\
This dependence on $\Gamma$ for low temperatures also extends to the linear transport coefficients defined in Eq.\ref{eq:lintranscoeff}, because they are proportional to $n_{\rm s}$, which is proportional to $\Gamma/\Delta$ for low temperatures. Therefore, the linear model is strongly dependent on the value of $\Gamma$, which can be strongly realization-dependent in a real experiment, making the model less reliable. As shown below in Sec.\ref{sec:gamma}, the nonlinear approach properly accounts for the effects of $\Gamma$.
\subsection{Nonlinear thermoelectricity}
\begin{figure*}[tb]
\centering
\begin{tikzpicture}
  \node[anchor=south west, inner sep=0] (img) at (0,0)
    {\includegraphics[width=\textwidth]{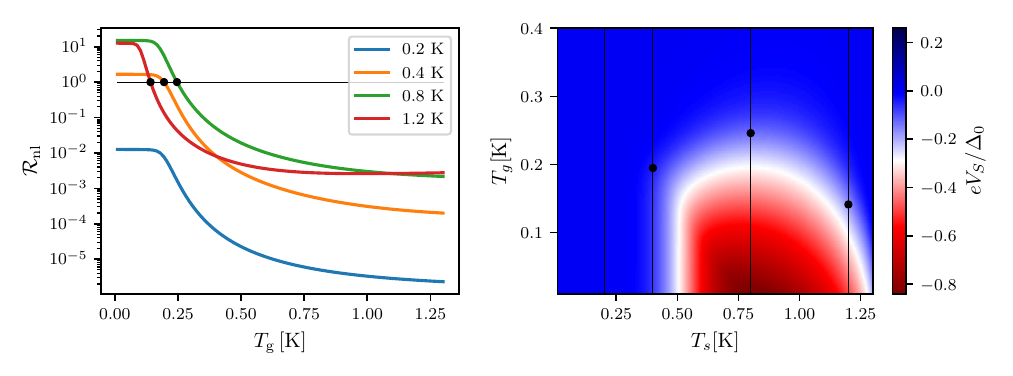}};
  \begin{scope}[x={(img.south east)}, y={(img.north west)}]
    % Coordinate system: (0,0) is bottom-left, (1,1) is top-right of image
    \node[anchor=north west, color=black] at (0.02,0.94) {\textbf{a)}};
    \node[anchor=north west, color=black] at (0.47,0.94) {\textbf{b)}};
  \end{scope}
\end{tikzpicture}
\caption{Testing our criterion for nonlinear thermoelectricity on the GIS junction. {\bf a)} Dependence of ratio $\mathcal{R}_{\rm nl}$ (defined in text) on $\Tg$ for different values of $\Ts$. The black line corresponds to $\mathcal{R}_{\rm nl}=1$, and it does not intercept $\Tg=0.2$ K. {\bf b)} Colormap representing $V_S(\Ts-\Tg)$. The black vertical lines are the paths plotted in a). The points corresponding to $\mathcal{R}_{\rm nl}=1$ delimit the area where $V_S \gtrsim 0.1 \Delta_0$ with good approximation, and the line at $\Tg=0.2$ K only encounters linear thermoelectricity as expected from a).}
\label{fig:nonlinthermoel}
\end{figure*}
Nonlinear thermoelectricity occurs in open-circuit superconducting junctions whenever a small potential difference $\delta V$ building up in the junction is not able to cancel the Peltier current $I_0$ induced by a small temperature difference $\delta T$. 
This happens when the superconductor is the hotter side, and therefore the relative shift between $n_{\rm s}(E)$ and the colder distribution $f(E-eV,T_{\rm g})$ induced by $\delta V$ does not appreciably modify the current $I_{\rm g\rightarrow s}\propto \int dE n_{\rm s}(E)f(E-eV,T_{\rm g})\sim f(\Delta-eV,T_{\rm g})$, that should cancel $I_0$. The uncompensated $I_0$ builds up charge on one side of the junction until the Fermi energy of the cold distribution reaches a peak of $n_{\rm s}(E)$, thus canceling the thermoelectric current. This is why $|V_S|\sim \Delta/e$ in this regime. Clearly, this effect cannot be described by linear modelling, as we need the full nonlinear shape of both $n_{\rm s}$ and $f$. \\
Although nonlinear thermoelectricity has been introduced in previous works \cite{Marchegiani2020, Bianco2024}, there is no proper criterion in the literature identifying the regime where it dominates over linear thermoelectricity.\\
Considering the intuitive mechanism, we can say that nonlinear thermoelectricity occurs when the linearized current variation induced by a potential equal to $\Delta/e$ is smaller than the linearized variation in current caused by a typical temperature variation on the colder side ($\sim T_{\rm g}$) 
\begin{equation}\label{eq:nonlincrit}
    \frac{\partial I}{\partial V}\frac{\Delta}{e}< \frac{\partial I}{\partial T} T_c \implies \frac{e}{1.764 k_B}\frac{\partial I/\partial T}{\partial I/\partial V}\equiv \mathcal{R}_{\rm nl}<1.
\end{equation}
We can confirm the validity of this criterion by numerically computing $V_S$ for the GIS junction and checking if the condition $\mathcal{R}_{\rm nl}>1$ actually delimits the area in which $V_S$ is comparable to $\Delta$. Figure \ref{fig:nonlinthermoel} shows that the points where $\mathcal{R}_{\rm nl}=1$ correctly delimit the area with $|V_S|\sim \Delta$. This criterion is quite easy to compute, allowing for a more straightforward understanding of nonlinear thermoelectricity in new, promising systems. 

\section{Nonlinear model and effects of nonlinearity}\label{sec:effects}
\subsection{Numerical solution of full model}
As the system of equations in Eq.\ref{eq:thermdyn} is not analytically solvable, we can only resort to numerics. The thermal dynamics equations 
describe the dependence of the junction temperatures on time $T_{\rm s,g}(t)$. \\
Therefore, we solve the initial value problem defined by the system of ordinary differential equations in Eq.\ref{eq:thermdyn} and two initial temperatures $T_{\rm s0}, T_{\rm g0}$ numerically. The solution of Eq.\ref{eq:thermdyn} requires repeated evaluation of the integrals in Eq. \ref{eq:thermsuper}, \ref{eq:epsc}, and \ref{eq:qgis}. All these integrals are improper because of the diverging $n_{\rm S}(E,\Ts)$ (Eq.\ref{eq:thermsuper}), and they therefore require relatively long computation times. A reasonable solution to this problem is sampling these integrals on a $(\Ts,\Tg)$ grid and building subsequent interpolation functions.
We computed on the $(\Ts,\Tg)$ grid the following quantities:
\begin{itemize}
    \item $S_{\rm s}(\Ts)$ (Eq. \ref{eq:thermsuper}),
    \item $Q_{\rm s-ph}(\Ts)$ (Eq. \ref{eq:epsc}),
    \item $Q_{\rm GIS}(\Ts,\Tg)$ (Eq. \ref{eq:qgis}),
    \item the Seebeck potential $V_S(\Ts,\Tg)$, computed from $I(V_S,\Ts,\Tg)=0$,
    \item the Peltier current, computed as $I(0,\Ts,\Tg)$,
    \item the differential resistance $R_D(V_S,\Ts,\Tg)$, computed as $dI(V,\Ts,\Tg)/dV|_{V_S}$.
\end{itemize}
This solution only works if the integrals exhibit sufficiently smooth behavior and if we are exploring temperature differences that are much larger than the grid spacing.\\
\subsection{New regimes}\label{sec:newreg}
\begin{figure*}[tb]
\centering
\begin{tikzpicture}
  \node[anchor=south west, inner sep=0] (img) at (0,0)
    {\includegraphics[width=\textwidth]{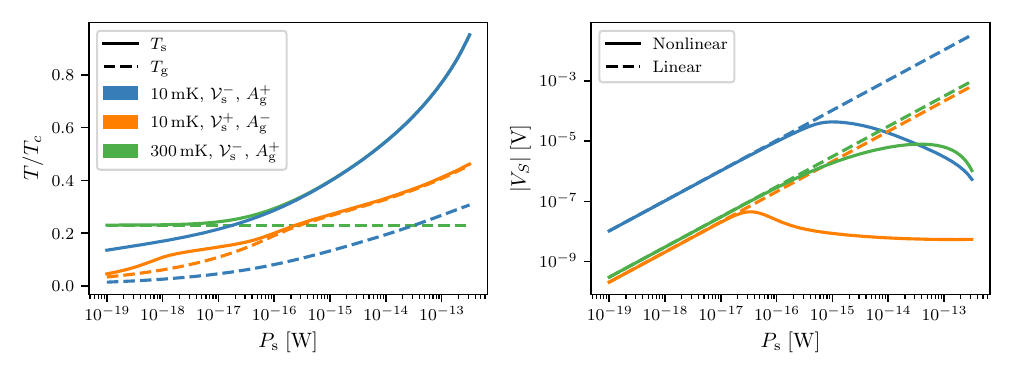}};
  \begin{scope}[x={(img.south east)}, y={(img.north west)}]
    % Coordinate system: (0,0) is bottom-left, (1,1) is top-right of image
    \node[anchor=north west, color=black] at (0.02,0.94) {\textbf{a)}};
    \node[anchor=north west, color=black] at (0.49,0.94) {\textbf{b)}};
  \end{scope}
\end{tikzpicture}
\caption{New regime of nonlinear model. a) Steady-state temperatures $\Ts$ and $\Tg$ dependence on power impinging on superconductor $P_{\rm s}$ for two different geometries (see main text in Sec. \ref{sec:newreg}) and bath temperatures $\Tb$. For lower bath temperatures, $\Tg$ increases with $P_{\rm s}$ because this increases dissipation. Geometry can destroy the temperature gradient $\Ts-\Tg$.  b) Dependence of $V_S$ on $P_{\rm s}$ for both the linear and the nonlinear model. $V_S$ deviates from the linear model for very low powers $P_{\rm s}\lesssim 10^{-17}\,{\rm W}$ even for more favorable geometries $(\mathcal{V}_{\rm s}^-,A_{\rm g}^+)$. $V_S$ from the nonlinear model shows a non-monotonous behavior due to the nonlinearity of the thermal equilibrium equations, which pins $\Ts$ to an almost constant value over a certain $P_{\rm s}$ and increases $\Tg$, reducing $V_S$. This is a new regime that cannot be described by linear modelling (see main text).}
\label{fig5}
\end{figure*}

The interplay between heat and charge transport introduces nontrivial behaviors for this relatively simple system. 
In Fig.\ref{fig5}a), we can notice how geometry can affect the formation of a thermal gradient across the junction. We represent the dependence of $\Ts$ and $\Tg$ on $P_{\rm s}$ for two very different geometries, one with $\mathcal{V}_s^-=5\cdot10^{-22}\,\mu {\text m}^3,\,A_{\rm j}^-=10^{-2} \,\mu{\text m}^2,\,A_{\rm g}^+=30 \,\mu{\text m}^2$, and one with $\mathcal{V}_s^+=10^{-19}\,\mu {\text m}^3,\,A_{\rm j}^-=0.5 \,\mu{\text m}^2,\,A_{\rm g}^+=1 \,\mu{\text m}^2$. The goal is to explore the most diverse regimes we can realistically achieve for this junction. While in the linear model a gradient always forms (see Eq. \ref{eqdeltaTlin}), the formation of a thermal gradient in the nonlinear model depends on nonlinear effects. For the first geometry, a thermal gradient forms for both values of $\Tb$, because heat flow across the junction is mainly limited by $Q_{\rm GIS}$ and not by $Q_{\rm g-ph}$. In this case, $\Tg$ does not need to rise much to dissipate the heat coming from the junction. In turn, the second geometry allows us to see what happens when $Q_{\rm g-ph}$ limits the heat flow. In this case, $\Tg$ rises to the point where $Q_{\rm GIS}$ is suppressed, because the rise in $Q_{\rm g-ph}$ is not sufficient to dissipate the heat. This destroys the thermal gradient across the junction, as we can see for $P_{\rm s}\gtrsim10^{-16}\,{\text W}$ for the second geometry in Fig. \ref{fig5}a). Therefore, our model can describe regimes where the thermal gradient does not form, even with a constant power impinging on the superconductor.\\
This effect is also shown in Figure \ref{fig5}b), which  represents the dependence of the Seebeck potential $V_S$ on the impinging power $P_{\rm s}$ for the linear and the nonlinear model for the same geometries and $\Tb$ values. Coherently with the behavior of $\Ts$ and $\Tg$ for the second geometry, we can see that the nonlinear value of $|V_S|$ deviates from its linear value and approaches very small values. 
Furthermore, we can notice that the behavior of the nonlinearly computed $V_S$ becomes non-monotonous. Surprisingly, this non-monotonicity extends to the other geometry, even in the presence of a thermal gradient.
This is due to two effects acting at the same time: the reduction of $\Delta$ with increasing $P_{\rm s}$, and the increase in $\Tg$ with increasing $P_{\rm s}$. These effects can be seen in Fig.\ref{fig5}a), where we represent the dependence of $\Ts$ and $\Tg$ on $P_{\rm s}$. At powers comparable with the departure from the linear model, $\Ts$ becomes larger than $0.4T_c$, which is the typical temperature at which $\Delta$ begins to get appreciably smaller. At the same time, $T_{\rm g}$ increases because a larger temperature gradient across the junction drives a larger heat flow that requires a larger $\Tg$ to be dissipated. The simultaneous presence of these two effects causes the counterintuitive decrease of $|V_S|$ with temperature.\\
This non-monotonous departure of $|V_S|$ from the linear behavior is therefore general and cannot be described by the linear model. The importance of this nonlinearity stems from the fact that it occurs for values of $P_{\rm s}$ relevant for applications \cite{Heikkilae2018,Giazotto2006}. 
\subsection{Effects of $\Gamma$ and quantitative comparison}\label{sec:gamma}
\begin{figure*}[tb]
\centering
\begin{tikzpicture}
  \node[anchor=south west, inner sep=0] (img) at (0,0)
    {\includegraphics[width=\textwidth]{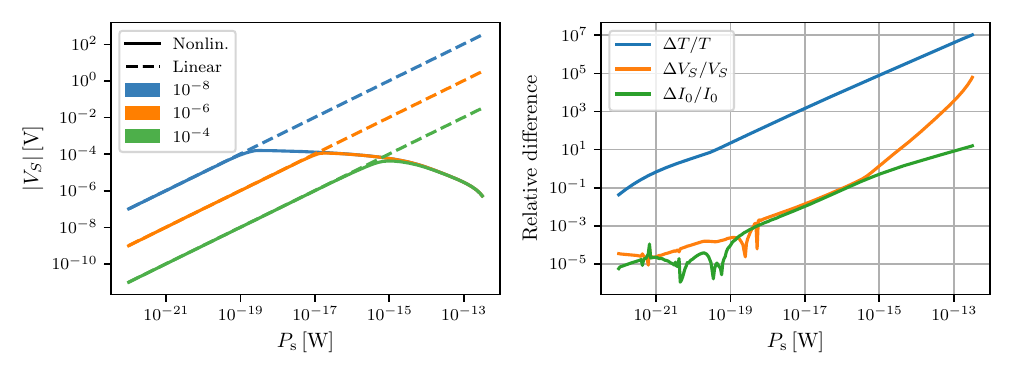}};
  \begin{scope}[x={(img.south east)}, y={(img.north west)}]
    \node[anchor=north west, color=black] at (0.02,0.94) {\textbf{a)}};
    \node[anchor=north west, color=black] at (0.5,0.94) {\textbf{b)}};
  \end{scope}
\end{tikzpicture}
\caption{Nonlinear model results. a) Dependence of steady-state $V_S$ on $P_{\rm s}$ for different values of $\Gamma$ for the linear and the nonlinear model. We observe that linear transport coefficients are proportional to $\Gamma$, and that decreasing $\Gamma$ reduces the value of $P_{\rm s}$ at which linearity is broken. Nonlinear model correctly describes onset of BCS quasiparticles. b) Relative difference $\Delta X/X=(X_{\rm lin}-X_{\rm nl})/X_{\rm nl}$ of observable quantities $X=\Ts,V_S,I_0$ between linear and nonlinear model. For all quantities, linearity is broken at very small impinging powers, with $\Ts$ breaking before $V_S$ and $I_0$ (explained in main text). Oscillations at low $P_{\rm s}$ are due to discretization effects.} 
\label{fig6}
\end{figure*}

As mentioned above in Sec.\ref{sec:subgap}, subgap states can have a substantial effect on transport, especially for low temperatures where BCS quasiparticle density is exponentially suppressed. They only produce linear thermoelectricity because they have an almost constant density of states $\Gamma/\Delta$. Since linear transport coefficients are computed for temperatures equal to $\Tb$, the value of $\Gamma$ strongly affects the results of the linear model. As we can see in Fig.\ref{fig6}a), values of $V_S$ computed with the linear and the nonlinear model begin to differ for smaller $P_{\rm s}$ when $\Gamma$ becomes smaller. Smaller values of $\Gamma$ induce an easier breaking of linearity, since there are fewer states available for linear thermoelectricity. \\
For larger $P_{\rm s}$, the effect of subgap states becomes less important because of the increase in $\Ts$ and the consequent formation of a larger BCS quasiparticle density. As shown in Fig.\ref{fig6}a), the linear model obviously fails to describe this effect. In contrast, the nonlinear model correctly describes the phenomenon, with $V_S$ computed at different $\Gamma$ converging for large $P_{\rm s}$. This resilience to the value of $\Gamma$ becomes more important whenever a model is used to describe realistic junctions for applications, where $\Gamma$ can also have larger values of $10^{-3}\text{--}10^{-2}\Delta_0$.\\
In Fig.\ref{fig6}b), we show the quantitative comparison of observable quantities $\Ts,\,V_S$ and $I_0$ between the linear and the nonlinear model for $\Tb=10\,{\rm mK}$. We represented the relative difference of the quantities (e.g. $(V_{S,{\rm lin}}-V_{S,{\rm nl}})/V_{S,{\rm nl}}$) to observe where the linear model is not able to describe the behavior of our system. While the linearity of $\Ts$ has already been represented in Fig. \ref{fig2}, we show it here again for comparison. We can assume that linearity is broken when the relative difference between a linear model-computed quantity and the corresponding nonlinear-computed quantity is more than $10\%$. We can see that, for all three quantities, linearity is broken at a very small $P_{\rm s}$, considering that the typical request for a bolometer measuring the Cosmic Microwave Background radiation is $P_{\rm s}\sim 10^{-12}\,{\rm W}$ \cite{DeLucia2024}. We can also see that, for $\Ts$, linearity is broken at very different $P_{\rm s}$ than the other quantities. In general, there is no reason for them to break linearity at the same $P_{\rm s}$. In this specific case, this difference is due to the cancellation of nonlinearities for $V_S$ and $I_0$ in the low-power limit. For low powers, the dominating contribution to thermal balance is given by $G_{\rm th}$ (see Eq.\ref{eq:lintranscoeff}.) If we perform a Taylor expansion for small $P_{\rm s}$, we can see that $\delta T/\delta P_{\rm s}\sim \partial Q_{\rm GIS}/\partial \Ts$, while $\delta I/\delta P_{\rm s}\sim \partial I/\partial \Ts/\partial Q_{\rm GIS}/\partial \Ts$. $\partial Q_{\rm GIS}/\partial \Ts$ and $\partial I/\partial \Ts$ are strongly nonlinear quantities, but they show the exact same behavior in $\Ts$. Therefore, they cancel out for $I_0$ and also for $V_S$, as their variations are proportional in the low-power limit.\\
In general, the linear model does not work in any reasonable regime to describe thermoelectricity in this type of superconducting hybrid junction.
\section{Conclusions}
The primary result obtained in this work is the assessment of the limits of linear modeling in describing thermoelectricity in a representative superconducting hybrid junction, as well as the development of a general nonlinear modeling framework. While the fact that a linear model does not work outside low-temperature or low-impinging power regimes is trivial, finding the limits of these regimes is far less trivial. We identified the limits of the linear model within the linear model itself, and we confirmed this by performing numerical simulations of the full system. We still consider the linear model to be useful for proof-of-concepts of thermoelectric mechanisms and devices \cite{Heikkilae2018,Chakraborty2018}. However, our work indicates that a nonlinear model is required for designing and realizing practical devices implemented in real experiments.\\
Additionally, we have identified the effect of superconductor subgap states in linear modeling and their strong influence on linearity breaking. We also discovered that the full system of equations can generate regimes that cannot be described by linear modelling alone, implying that geometry can affect the formation of a thermal gradient across the junction even with a relatively large power impinging on one side.\\
For the first time, we have identified a relatively simple and general criterion that marks the onset of nonlinear thermoelectricity, which can facilitate a more intuitive understanding of this phenomenon in various systems, thereby promoting the development of applications that utilize this interesting physical phenomenon.\\
In general, we believe that our work contributes to a deeper theoretical understanding of how thermoelectricity in superconducting hybrid tunnel junctions works in the nonlinear regime. This can expand their use cases for applications different than thermometry, possibly including particle and radiation detection \cite{Heikkilae2018,Paolucci2023} and cryogenic heat engines \cite{Germanese2022}. 
\begin{acknowledgments}
The authors wish to thank T.T. Heikkil{\"a} for fruitful discussion. L.L. also wants to thank G. Clemente for helpful discussion. The  Italian Ministry of University and Research funded the work of L.L. and F.P. under the call PRIN2022 (Financed by the European Union – Next Generation EU) project EQUATE (Grant No. 2022Z7RHRS).
\end{acknowledgments}


\begin{thebibliography}{10}

\bibitem{Shannon1948}
C.~E. Shannon.
\newblock A mathematical theory of communication.
\newblock {\em Bell System Technical Journal}, 27(3):379--423, July 1948.

\bibitem{Raussendorf2001}
Robert Raussendorf and Hans~J. Briegel.
\newblock A one-way quantum computer.
\newblock {\em Physical Review Letters}, 86(22):5188--5191, May 2001.

\bibitem{Koch2007}
Jens Koch, Terri~M. Yu, Jay Gambetta, A.~A. Houck, D.~I. Schuster, J.~Majer,
  Alexandre Blais, M.~H. Devoret, S.~M. Girvin, and R.~J. Schoelkopf.
\newblock Charge-insensitive qubit design derived from the cooper pair box.
\newblock {\em Physical Review A}, 76(4):042319, October 2007.

\bibitem{Valenzuela2006}
Sergio~O. Valenzuela, William~D. Oliver, David~M. Berns, Karl~K. Berggren,
  Leonid~S. Levitov, and Terry~P. Orlando.
\newblock Microwave-induced cooling of a superconducting qubit.
\newblock {\em Science}, 314(5805):1589--1592, December 2006.

\bibitem{Kjaergaard2020}
Morten Kjaergaard, Mollie~E. Schwartz, Jochen Braumüller, Philip Krantz, Joel
  I.-J. Wang, Simon Gustavsson, and William~D. Oliver.
\newblock Superconducting qubits: Current state of play.
\newblock {\em Annual Review of Condensed Matter Physics}, 11(1):369--395,
  March 2020.

\bibitem{Burkard2023}
Guido Burkard, Thaddeus~D. Ladd, Andrew Pan, John~M. Nichol, and Jason~R.
  Petta.
\newblock Semiconductor spin qubits.
\newblock {\em Reviews of Modern Physics}, 95(2):025003, June 2023.

\bibitem{Irwin1995}
K.~D. Irwin.
\newblock An application of electrothermal feedback for high resolution
  cryogenic particle detection.
\newblock {\em Applied Physics Letters}, 66(15):1998--2000, April 1995.

\bibitem{Ade2015}
P.~A.~R. Ade~et al.
\newblock bicep2/keck array. iv. optical characterization and performance of
  the bicep2 and keck array experiments.
\newblock {\em The Astrophysical Journal}, 806(2):206, June 2015.

\bibitem{Crowley2018}
Kevin~T. Crowley~et al.
\newblock Advanced actpol tes device parameters and noise performance in
  fielded arrays.
\newblock {\em Journal of Low Temperature Physics}, 193(3–4):328--336, July
  2018.

\bibitem{Mairs2021}
Steve Mairs~et al.
\newblock A decade of scuba-2: A comprehensive guide to calibrating 450 $\mu$m and
  850 $\mu$m continuum data at the jcmt.
\newblock {\em The Astronomical Journal}, 162(5):191, October 2021.

\bibitem{Giazotto2015}
F.~Giazotto, P.~Solinas, A.~Braggio, and F. S. Bergeret.
\newblock Ferromagnetic-insulator-based superconducting junctions as sensitive
  electron thermometers.
\newblock {\em Physical Review Applied}, 4(4):044016, October 2015.

\bibitem{Heikkilae2018}
T.~T. Heikkilä, R.~Ojajärvi, I.~J. Maasilta, E.~Strambini, F.~Giazotto, and
  F.~S. Bergeret.
\newblock Thermoelectric radiation detector based on superconductor-ferromagnet
  systems.
\newblock {\em Physical Review Applied}, 10(3):034053, September 2018.

\bibitem{Schwab2000}
K.~Schwab, E.~A. Henriksen, J.~M. Worlock, and M.~L. Roukes.
\newblock Measurement of the quantum of thermal conductance.
\newblock {\em Nature}, 404(6781):974--977, April 2000.

\bibitem{Jezouin2013}
S.~Jezouin, F.~D. Parmentier, A.~Anthore, U.~Gennser, A.~Cavanna, Y.~Jin, and
  F.~Pierre.
\newblock Quantum limit of heat flow across a single electronic channel.
\newblock {\em Science}, 342(6158):601--604, November 2013.

\bibitem{Giazotto2012}
Francesco Giazotto and María~José Martínez-Pérez.
\newblock The josephson heat interferometer.
\newblock {\em Nature}, 492(7429):401--405, December 2012.

\bibitem{JoseMartinezPerez2014}
Maria José Martínez-Pérez and Francesco Giazotto.
\newblock A quantum diffractor for thermal flux.
\newblock {\em Nature Communications}, 5(1), April 2014.

\bibitem{Fornieri2017}
Antonio Fornieri and Francesco Giazotto.
\newblock Towards phase-coherent caloritronics in superconducting circuits.
\newblock {\em Nature Nanotechnology}, 12(10):944--952, October 2017.

\bibitem{Chang2006}
C.~W. Chang, D.~Okawa, A.~Majumdar, and A.~Zettl.
\newblock Solid-state thermal rectifier.
\newblock {\em Science}, 314(5802):1121--1124, November 2006.

\bibitem{MartinezPerez2015}
Maria~José Martínez-Pérez, Antonio Fornieri, and Francesco Giazotto.
\newblock Rectification of electronic heat current by a hybrid thermal diode.
\newblock {\em Nature Nanotechnology}, 10(4):303--307, February 2015.

\bibitem{Timossi2018}
Giuliano~Francesco Timossi, Antonio Fornieri, Federico Paolucci, Claudio
  Puglia, and Francesco Giazotto.
\newblock Phase-tunable josephson thermal router.
\newblock {\em Nano Letters}, 18(3):1764--1769, February 2018.

\bibitem{Strambini2014}
E.~Strambini, F.~S. Bergeret, and F.~Giazotto.
\newblock Proximity nanovalve with large phase-tunable thermal conductance.
\newblock {\em Applied Physics Letters}, 105(8), August 2014.

\bibitem{Paolucci2017}
F.~Paolucci, G.~Marchegiani, E.~Strambini, and F.~Giazotto.
\newblock Phase-tunable temperature amplifier.
\newblock {\em EPL (Europhysics Letters)}, 118(6):68004, June 2017.

\bibitem{Paolucci2018}
Federico Paolucci, Giampiero Marchegiani, Elia Strambini, and Francesco
  Giazotto.
\newblock Phase-tunable thermal logic: Computation with heat.
\newblock {\em Physical Review Applied}, 10(2):024003, August 2018.

\bibitem{Ligato2022}
Nadia Ligato, Federico Paolucci, Elia Strambini, and Francesco Giazotto.
\newblock Thermal superconducting quantum interference proximity transistor.
\newblock {\em Nature Physics}, 18(6):627--632, April 2022.

\bibitem{Giazotto2006}
Francesco Giazotto, Tero~T. Heikkilä, Arttu Luukanen, Alexander~M. Savin, and
  Jukka~P. Pekola.
\newblock Opportunities for mesoscopics in thermometry and refrigeration:
  Physics and applications.
\newblock {\em Reviews of Modern Physics}, 78(1):217--274, March 2006.

\bibitem{Pirro2017}
S.~Pirro and P.~Mauskopf.
\newblock Advances in bolometer technology for fundamental physics.
\newblock {\em Annual Review of Nuclear and Particle Science}, 67(1):161--181,
  October 2017.

\bibitem{DeLucia2024}
Mario De~Lucia, Paolo Dal~Bo, Eugenia Di~Giorgi, Tommaso Lari, Claudio Puglia,
  and Federico Paolucci.
\newblock Transition edge sensors: Physics and applications.
\newblock {\em Instruments}, 8(4):47, October 2024.

\bibitem{Mao2020}
Jun Mao, Gang Chen, and Zhifeng Ren.
\newblock Thermoelectric cooling materials.
\newblock {\em Nature Materials}, 20(4):454--461, December 2020.

\bibitem{Chung2000}
Duck-Young Chung, Tim Hogan, Paul Brazis, Melissa Rocci-Lane, Carl Kannewurf,
  Marina Bastea, Ctirad Uher, and Mercouri~G. Kanatzidis.
\newblock CsBi$_4$Te$_6$ : A high-performance thermoelectric material for
  low-temperature applications.
\newblock {\em Science}, 287(5455):1024--1027, February 2000.

\bibitem{Sidorenko2019}
N.~A. Sidorenko and Z.~M. Dashevsky.
\newblock Cryogenic thermoelectric cooler for operating temperatures below 90
  k.
\newblock {\em Semiconductors}, 53(6):752--755, June 2019.

\bibitem{Grosso2014}
Giuseppe Grosso.
\newblock {\em Solid state physics}.
\newblock Elsevier, Waltham, Massachusetts, second edition (online-ausg.)
  edition, 2014.
\newblock Description based on online resource; title from PDF title page
  (ebrary, viewed November 19, 2013).

\bibitem{Hicks1993}
L.~D. Hicks and M.~S. Dresselhaus.
\newblock Effect of quantum-well structures on the thermoelectric figure of
  merit.
\newblock {\em Physical Review B}, 47(19):12727--12731, May 1993.

\bibitem{Hicks1993a}
L.~D. Hicks and M.~S. Dresselhaus.
\newblock Thermoelectric figure of merit of a one-dimensional conductor.
\newblock {\em Physical Review B}, 47(24):16631--16634, June 1993.

\bibitem{Sothmann2014}
Björn Sothmann, Rafael Sánchez, and Andrew~N Jordan.
\newblock Thermoelectric energy harvesting with quantum dots.
\newblock {\em Nanotechnology}, 26(3):032001, December 2014.

\bibitem{Jaliel2019}
G.~Jaliel, R. K. Puddy, R.~Sánchez, A. N. Jordan, B.~Sothmann, I.~Farrer,
  J. P. Griffiths, D. A. Ritchie, and C. G. Smith.
\newblock Experimental realization of a quantum dot energy harvester.
\newblock {\em Physical Review Letters}, 123(11):117701, September 2019.

\bibitem{Urban2015}
Jeffrey~J. Urban.
\newblock Prospects for thermoelectricity in quantum dot hybrid arrays.
\newblock {\em Nature Nanotechnology}, 10(12):997--1001, December 2015.

\bibitem{Ozaeta2014}
A.~Ozaeta, P.~Virtanen, F. S. Bergeret, and T. T. Heikkilä.
\newblock Predicted very large thermoelectric effect in
  ferromagnet-superconductor junctions in the presence of a spin-splitting
  magnetic field.
\newblock {\em Physical Review Letters}, 112(5):057001, February 2014.

\bibitem{Marchegiani2020}
G.~Marchegiani, A.~Braggio, and F.~Giazotto.
\newblock Nonlinear thermoelectricity with electron-hole symmetric systems.
\newblock {\em Physical Review Letters}, 124(10):106801, March 2020.

\bibitem{Melton1980}
Robert~G. Melton, James~L. Paterson, and S.~B. Kaplan.
\newblock Superconducting tunnel-junction refrigerator.
\newblock {\em Physical Review B}, 21(5):1858--1867, March 1980.

\bibitem{Nahum1994}
M.~Nahum, T.~M. Eiles, and John~M. Martinis.
\newblock Electronic microrefrigerator based on a
  normal-insulator-superconductor tunnel junction.
\newblock {\em Applied Physics Letters}, 65(24):3123--3125, December 1994.

\bibitem{Leivo1996}
M.~M. Leivo, J.~P. Pekola, and D.~V. Averin.
\newblock Efficient peltier refrigeration by a pair of normal
  metal/insulator/superconductor junctions.
\newblock {\em Applied Physics Letters}, 68(14):1996--1998, April 1996.

\bibitem{Tarasov2020}
Mikhail~A. Tarasov, Aleksandra~A. Gunbina, Sumedh Mahashabde, Renat~A. Yusupov,
  Artem~M. Chekushkin, Daria~V. Nagirnaya, Valerian~S. Edelman, Grigory~V.
  Yakopov, and Vyacheslav~F. Vdovin.
\newblock Arrays of annular antennas with sinis bolometers.
\newblock {\em IEEE Transactions on Applied Superconductivity}, 30(3):1--6,
  April 2020.

\bibitem{Manninen1999}
A.~J. Manninen, J.~K. Suoknuuti, M.~M. Leivo, and J.~P. Pekola.
\newblock Cooling of a superconductor by quasiparticle tunneling.
\newblock {\em Applied Physics Letters}, 74(20):3020--3022, May 1999.

\bibitem{Kolenda2016}
S.~Kolenda, M. J. Wolf, and D.~Beckmann.
\newblock Observation of thermoelectric currents in high-field
  superconductor-ferromagnet tunnel junctions.
\newblock {\em Physical Review Letters}, 116(9):097001, March 2016.

\bibitem{Germanese2022}
Gaia Germanese, Federico Paolucci, Giampiero Marchegiani, Alessandro Braggio,
  and Francesco Giazotto.
\newblock Bipolar thermoelectric josephson engine.
\newblock {\em Nature Nanotechnology}, 17(10):1084--1090, September 2022.

\bibitem{Germanese2023}
Gaia Germanese, Federico Paolucci, Giampiero Marchegiani, Alessandro Braggio,
  and Francesco Giazotto.
\newblock Phase control of bipolar thermoelectricity in josephson tunnel
  junctions.
\newblock {\em Physical Review Applied}, 19(1):014074, January 2023.

\bibitem{Vischi2020}
Francesco Vischi, Matteo Carrega, Alessandro Braggio, Federico Paolucci,
  Federica Bianco, Stefano Roddaro, and Francesco Giazotto.
\newblock Electron cooling with graphene-insulator-superconductor tunnel
  junctions for applications in fast bolometry.
\newblock {\em Physical Review Applied}, 13(5):054006, May 2020.

\bibitem{Bianco2024}
Federica Bianco, Ding Zhang, and Federico Paolucci.
\newblock Coexistence of linear and non-linear thermoelectricity in
  graphene-superconductor tunnel junctions.
\newblock {\em Journal of Applied Physics}, 136(15), October 2024.

\bibitem{Mather1982}
John~C. Mather.
\newblock Bolometer noise: nonequilibrium theory.
\newblock {\em Applied Optics}, 21(6):1125, March 1982.

\bibitem{Richards1994}
P.~L. Richards.
\newblock Bolometers for infrared and millimeter waves.
\newblock {\em Journal of Applied Physics}, 76(1):1--24, July 1994.

\bibitem{Golubev2001}
Dmitri Golubev and Leonid Kuzmin.
\newblock Nonequilibrium theory of a hot-electron bolometer with normal
  metal-insulator-superconductor tunnel junction.
\newblock {\em Journal of Applied Physics}, 89(11):6464--6472, June 2001.

\bibitem{Chakraborty2018}
Subrata Chakraborty and Tero~T. Heikkilä.
\newblock Thermoelectric radiation detector based on a
  superconductor-ferromagnet junction: Calorimetric regime.
\newblock {\em Journal of Applied Physics}, 124(12), September 2018.

\bibitem{Whitney2013}
Robert~S. Whitney.
\newblock Nonlinear thermoelectricity in point contacts at pinch off: A
  catastrophe aids cooling.
\newblock {\em Physical Review B}, 88(6):064302, August 2013.

\bibitem{Talbo2017}
Vincent Talbo, Jérôme Saint-Martin, Sylvie Retailleau, and Philippe Dollfus.
\newblock Non-linear effects and thermoelectric efficiency of quantum dot-based
  single-electron transistors.
\newblock {\em Scientific Reports}, 7(1), November 2017.

\bibitem{Paolucci2023}
Federico Paolucci, Gaia Germanese, Alessandro Braggio, and Francesco Giazotto.
\newblock A highly sensitive broadband superconducting thermoelectric
  single-photon detector.
\newblock {\em Applied Physics Letters}, 122(17), April 2023.

\bibitem{Dynes1984}
R.~C. Dynes, J.~P. Garno, G.~B. Hertel, and T.~P. Orlando.
\newblock Tunneling study of superconductivity near the metal-insulator
  transition.
\newblock {\em Physical Review Letters}, 53(25):2437--2440, December 1984.

\bibitem{CastroNeto2009}
A.~H. Castro~Neto, F.~Guinea, N.~M.~R. Peres, K.~S. Novoselov, and A.~K. Geim.
\newblock The electronic properties of graphene.
\newblock {\em Reviews of Modern Physics}, 81(1):109--162, January 2009.

\bibitem{Timofeev2009}
A.~V. Timofeev, C.~Pascual García, N.~B. Kopnin, A.~M. Savin, M.~Meschke,
  F.~Giazotto, and J.~P. Pekola.
\newblock Recombination-limited energy relaxation in a
  bardeen-cooper-schrieffer superconductor.
\newblock {\em Physical Review Letters}, 102(1):017003, January 2009.

\bibitem{Chen2012}
Wei Chen and Aashish~A. Clerk.
\newblock Electron-phonon mediated heat flow in disordered graphene.
\newblock {\em Physical Review B}, 86(12):125443, September 2012.

\bibitem{Zihlmann2019}
Simon Zihlmann, Péter Makk, Sebastián Castilla, Jörg Gramich, Kishan
  Thodkar, Sabina Caneva, Ruizhi Wang, Stephan Hofmann, and Christian
  Schönenberger.
\newblock Nonequilibrium properties of graphene probed by superconducting
  tunnel spectroscopy.
\newblock {\em Physical Review B}, 99(7):075419, February 2019.

\bibitem{Onsager1931}
Lars Onsager.
\newblock Reciprocal relations in irreversible processes. ii.
\newblock {\em Physical Review}, 38(12):2265--2279, December 1931.

\end{thebibliography}
\end{document}